\documentclass[12pt]{article}
\usepackage{setspace}
\usepackage{graphicx}
\usepackage[margin=1in]{geometry}
\usepackage{amsmath}
\usepackage{multicol}
\usepackage{titlesec}
\usepackage{abstract}
\usepackage{float}

\titleformat{\section}
  {\normalfont\fontsize{11}{15}\bfseries}{\thesection}{1em}{}
\titleformat{\subsection}
  {\normalfont\fontsize{11}{15}\bfseries}{\thesubsection}{1em}{}
 
\titleformat{\subsubsection}
  {\normalfont\fontsize{11}{15}\bfseries}{\thesubsubsection}{1em}{}
  
\graphicspath{ {images/} }
\usepackage{amsmath} 
\usepackage[style=phys]{biblatex}
\usepackage{caption}

\renewcommand{\thesection}{\Roman{section}} 
\DeclareFieldFormat{titlecase}{#1}

\bibliography{lib.bib}
\title{\vspace{-2cm}\fontsize{14}{16} \selectfont \textbf{Spatial Mapping of Electrostatics and Dynamics across 2D Heterostructures}}

\author{\fontsize{12}{15} \selectfont Akshay A. Murthy$^{1,2}$, Stephanie M. Ribet$^{1}$, Teodor K. Stanev$^3$, Pufan Liu$^1$,\\
\fontsize{12}{15} \selectfont Kenji Watanabe$^4$, Takashi Taniguchi$^5$, Nathaniel P. Stern$^3$, Roberto dos Reis$^{1,6,*}$,\\
\fontsize{12}{15} \selectfont Vinayak P. Dravid$^{1,2,6,*}$}

\date{%
    \fontsize{10}{15} \selectfont
    $^1$Department of Materials Science and Engineering, Northwestern University, Evanston, IL, USA\\%
    $^2$International Institute of Nanotechnology, Northwestern University, Evanston, IL, USA\\
    $^3$Department of Physics and Astronomy, Northwestern University, Evanston, IL, USA\\
    $^4$Research Center for Functional Materials, National Institute for Materials Science, 1-1 Namiki, Tsukuba 305-0044, Japan\\
    $^5$International Center for Materials Nanoarchitectonics, National Institute for Materials Science, 1-1 Namiki, Tsukuba 305-0044, Japan\\
    $^6$The NU\textit{ANCE} Center, Northwestern University, Evanston, IL, USA\\~\\
    $^*$Corresponding Authors:\\
    Roberto dos Reis: \underline{roberto.reis@northwestern.edu}\\
    Vinayak P. Dravid: \underline{v-dravid@northwestern.edu}
}

\begin{document}
\singlespacing
\maketitle

\begin{abstract}
\noindent
\textit{In situ} electron microscopy is a key tool for understanding the mechanisms driving novel phenomena in 2D structures. Unfortunately, due to various practical challenges, technologically relevant 2D heterostructures prove challenging to address with electron microscopy. Here, we use the differential phase contrast imaging technique to build a methodology for probing local electrostatic fields during electrical operation with nanoscale precision in such materials. We find that by combining a traditional DPC setup with a high pass filter, we can largely eliminate electric fluctuations emanating from short-range atomic potentials. With this method, \textit{a priori} electric field expectations can be directly compared with experimentally derived values to readily identify inhomogeneities and potentially problematic regions. We use this platform to analyze the electric field and charge density distribution across layers of hBN and MoS$_2$. 

\end{abstract}
\begin{multicols}{2}
\section{INTRODUCTION}
The diverse class of two-dimensional (2D) materials including graphene, hexagonal boron nitride (hBN), and transition metal dichalcogenides (TMDs), have produced a wide variety of captivating and technologically relevant features for next generation optoelectronic architectures and quantum information science.\cite{RN1756, RN1757, RN1758} These include unique memory devices,\cite{RN1748, RN1749} Josephson junctions for solid state qubit architectures,\cite{RN1809,RN1810} photonic structures,\cite{RN1753, RN1754, RN1755} and electronic devices.\cite{RN472, RN341, RN1760} Although charge transport dynamics present at various 2D or contact interfaces contribute significantly to device performance in such structures, the atomistic features giving rise to macroscopic properties often remains unclear. This makes it difficult to identify the optimal methods for improving performance and reliability. \textit{In situ} electrical biasing scanning/transmission electron microscopy (S/TEM) is a particularly attractive method for probing property-performance relationships at these atomic length scales. This approach involves stimulating the sample of interest with an external electrical field while simultaneously probing it with a high energy electron beam. It is thus possible to directly ascertain the evolution of the atomic structure with nanoscale spatial resolution, such as electromigration of atoms in MoS$_{2}$.\cite{RN1596} 

Beyond providing structural information, the multimodal nature of STEM makes it possible to detect emergent phenomena. For instance, electric and/or magnetic fields within the sample partially deflect the incident electron probe leading to shifts in the intensity center of mass (\textit{I}\textsubscript{COM}) in the diffraction patterns. The result of this beam-specimen interaction can be measured at the sample plane through methods including differential phase contrast (DPC).\cite{RN1736,RN1737, RN1738} Traditional DPC makes use of the idea that the difference in measured diffraction intensities between detector regions diametrically opposite one another is directly proportional to the deflection of the electron probe along the common detector axis.\cite{RN1734,RN1735,RN1752, RN1746} Similarly, first moment (FM)–STEM configuration calculates \textit{I}\textsubscript{COM} directly by summing up the measured intensity at detector pixel. High-speed direct detection cameras \cite{RN1811, RN1743, RN1744} have opened the possibility to capture the scattered electron probe at each pixel in a real space image with roughly 10$^{-3}$ second dwell times \cite{RN1768} and enabled both DPC methods.\cite{RN1795}

Here we investigate 2D layers that can be simultaneously biased electronically while being probed with an electron beam. Using DPC to calculate variations in the center of mass of the electron probe, we are able to map the resultant electrostatic fields and charge densities that arise in 2D layers when an external bias is applied. By combining a traditional DPC setup with a high pass filter, we are able to selectively isolate the low spatial frequency information associated with these long-range potentials. This method paves the way for exploring a wide variety of 2D memory, quantum, and optoelectronic heterostructure devices.

\begin{figure*}
\includegraphics[width=\linewidth]{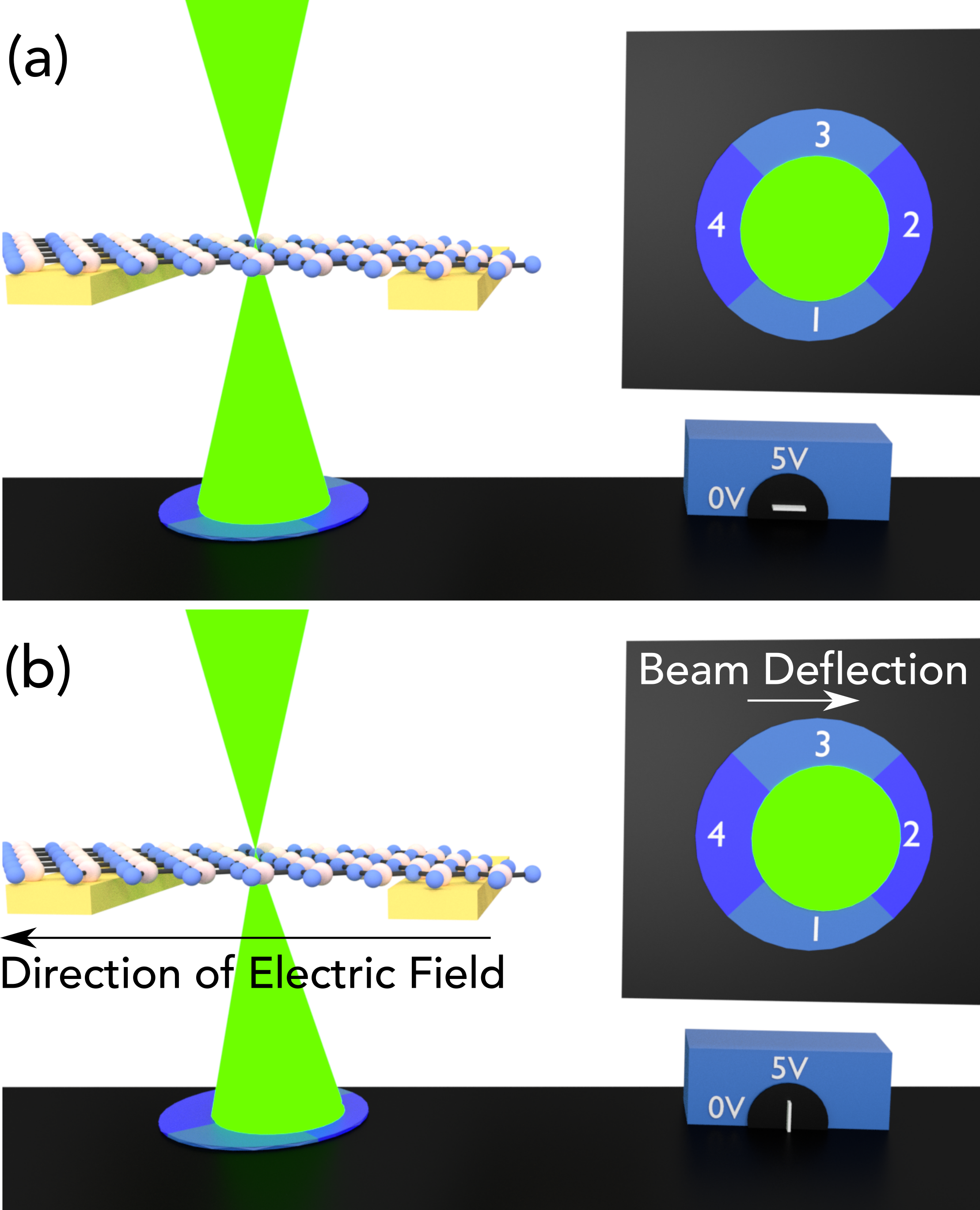}
\caption{Schematic of STEM electrical field measurement setup (a) In this approach, an electrical potential is applied across a 2D material system leading to a deflection in the center of mass of the electron diffraction pattern. (b) This momentum transfer experienced by the incident electrons in the detector plane is used to calculate the electric field vector at each probe position.}
\label{fgr:schematic}
\end{figure*}

\section{SAMPLE PREPARATION}
Although fabricating TEM devices of 2D materials and heterostructures is typically a challenging task, it is possible to construct such structures using a process described previously.\cite{RN1596} In this setup, 75 nm thick gold electrodes are patterned using electron beam lithography on a SiO$_{2}$/Si (300 nm thickness) substrate. This substrate is then spin-coated with a polycarbonate solution (5\% polycarbonate to 95\% chloroform w/v) at 2000 RPM for 60 seconds. After this coating is baked at 120$^{\circ}$C for 1 minute, the substrate is placed in DI water until the polymer containing the gold electrodes naturally released from the substrate. Once the polymer film had dried, it is carefully placed on two strips of PDMS gel attached to a glass slide and suspended roughly 1 mm above the glass slide. Using a micromanipulator stage, the gold electrodes are aligned over a TEM window (TEMwindows.com) and brought into contact with the TEM grid. This assembly is then gradually heated up to 150$^{\circ}$C for a few minutes before the sample is baked at 170$^{\circ}$C for 15 minutes. The structure is then placed in a chloroform bath for 6 hours to dissolve the polycarbonate film and moved to an IPA bath afterwards for a few minutes to remove chloroform residue.

Once a thin layer of mechanically exfoliated hBN (Figure S1) or a mechanically transferred MoS$_{2}$/hBN heterostructure is identified on a SiO$_{2}$/Si substrate, a similar process is then repeated to move the layers from the substrate to the TEM grid. The resultant sample is left overnight in chloroform and moved to an IPA bath afterwards. The sample is then annealed in a 95\% Ar/5\% H$_2$ mixture at 200$^{\circ}$C for 2 hours to remove residue prior to STEM imaging. Finally, gold wires are used to connect the electrode pads to the contact points on the Nanofactory\textsuperscript{\textregistered} TEM holder using indium as a bonding agent. Following fabrication of the sample, Raman and photoluminescence (PL) spectra are taken to confirm the presence of suspended layers of MoS$_{2}$ and hBN (Figure S2). The current-voltage characteristics that arise when a voltage is applied across this sample is provided in (Figure S2) demonstrating the ability to apply an external field across the sample. 

\section{EXPERIMENTAL}

S/TEM images of 2D layers are acquired using a JEOL ARM 300F Grand ARM S/TEM operated at 300 kV. The camera length is set to 8 cm and the condenser aperture is chosen to result in a convergence semi-angle of 30 mrad with beam current ~20pA to minimize beam-induced damage. Application of a high convergent angle (30 mrad) is intentionally chosen to optimize the measurement of a constant field in the sample that causes a uniform shift of the unscattered probe. The asymmetry in the measured pattern creates a difference signal that is proportional to the field strength. 4D STEM data set containing over 40,000 convergent beam electron diffraction (CBED) patterns are acquired in a 680 x 80 mesh (pixel size of 5.96Å) across the sample using Gatan® K3-IS in counting mode. A frame rate of 285 fps (0.0035s per diffraction) is chosen to prioritize acquisition of high resolution CBED (1024 x 1024 pixels) while maintaining sufficient sample stability under external bias.

\section{THEORY: Deducing In-Plane Electric Fields from Intensity Variations in CBED Patterns}

Intensity variations in the CBED pattern can be related to in-plane electric fields through the relations discussed in this section. When an electric field is present in the sample plane, the incident beam of electrons experience a Coulombic force leading to a deflection in the CBED pattern parallel to the detector plane. As discussed, DPC and FM-STEM offer methods to measure these changes in the \textit{I}\textsubscript{COM} of the CBED pattern as a function of position. By rearranging DeBroglie's wave equation and normalizing the overall electron probe intensity, the \textit{I}\textsubscript{COM} values calculated at each probe position can be related to the expectation value of the momentum transfer, $\left< p_{\bot }\right>$ (Eq. \ref{eqn:mom_transfer}). In this equation, \textit{h} represents Planck's constant and $\lambda$ represents electron wavelength.

\begin{equation}
   \left< p_{\bot }\right>= h \frac{\sin(I_{COM})}{\lambda} \label{eqn:mom_transfer}
\end{equation}

Further, \citeauthor{RN1751} showed that based on Ehrenfest's theorem, the measured in-plane electric field is given by Eq. \ref{eqn:e-field}.\cite{RN1751} In this equation, $v$ represents the velocity of electrons carrying a charge $e$ that interact with the sample for a time equal to $v$/$z$, where $z$ represents the thickness of the sample.\cite{RN1751, SEKI2017258} 

\begin{equation}
   E_{\bot } = \left< p_{\bot }\right> \frac{v}{ez} \label{eqn:e-field}
\end{equation}

Because this equation assumes that the intensity of the electron probe does not vary with $z$, this relation requires specimen thicknesses to be restricted to values less than 5 nm to avoid beam broadening effects.\cite{RN1751, MULLERCASPARY201762} In the case of thicker samples, sources of diffraction contrast can also contribute to probe deflection and introduce imaging artifacts.\cite{MACLAREN201557}

\begin{figure*}
\includegraphics[width=\linewidth]{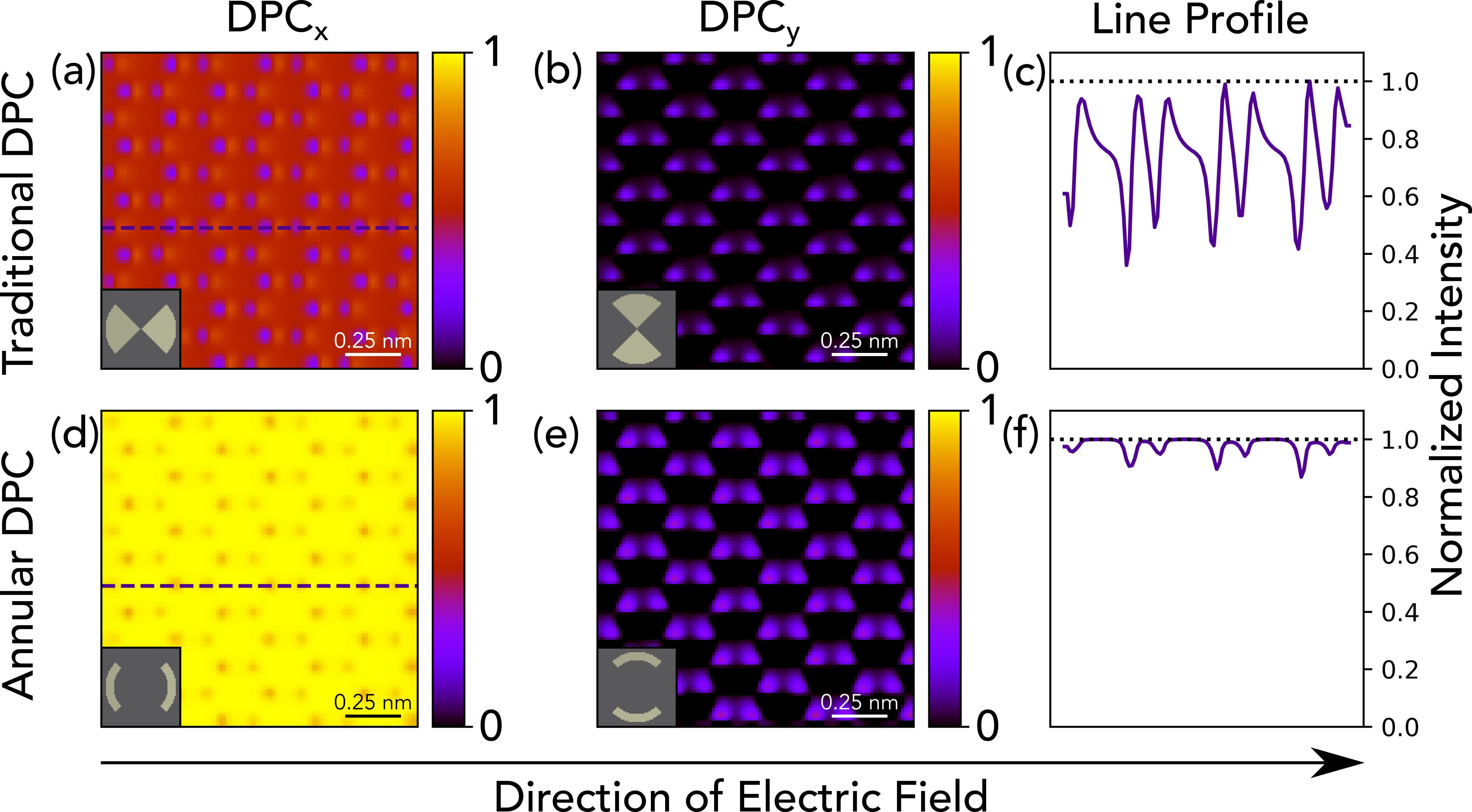}
\caption{DPC simulations for interrogating long-range electric fields (a-b) Simulated DPC images taken from hBN structure subjected to an varying electrical potential in the vertical direction detailed in Figure S4. The detector configurations are provided in the insets. (c) An intensity line profile taken from (a). (d-e) Simulated annular DPC images taken from same hBN structure with detector configurations are provided in the insets. (f) An intensity line profile taken from (c). From these line profiles, the simulated data suggests that intensity deviations from a uniform contrast profile resulting from atomic electric fields are significantly minimized in the annular geometry.}
\label{fgr:simulation}
\end{figure*}

Furthermore, as discussed by \citeauthor{10.1093/jmicro/dfx123}, it is apparent that low spatial frequency information associated with long-range fields leads to a uniform displacement of the CBED pattern.\cite{10.1093/jmicro/dfx123} Meanwhile, high spatial frequency information associated with short-range fields leads to a redistribution in intensity within the CBED pattern.\cite{10.1093/jmicro/dfx123} As such, one method to isolate information associated with long-range fields is through the use of an annular DPC detector.\cite{104427} Namely, by filtering out low frequency signal with an inner collection angle approaching the convergence semi-angle of the CBED pattern (Figure S5), the redistribution in signal associated with short-range potentials can be largely diminished. Simulated differential phase contrast images of hBN subjected to an external electric field demonstrate this aspect (Figure \ref{fgr:simulation}). In the case of a traditional DPC image (Figure \ref{fgr:simulation}a-b), local atomic electrostatic potentials and long-range external potentials can both contribute to the measured signal. In annular DPC images, however, (parameters listed in Methods) the contribution of local potentials can be selectively diminished (line profiles provided in Figure \ref{fgr:simulation}c \& f), which makes this method potentially very useful for mapping long-range electric fields.

\section{RESULTS AND DISCUSSION}

\begin{figure*}
\includegraphics[width=140mm]{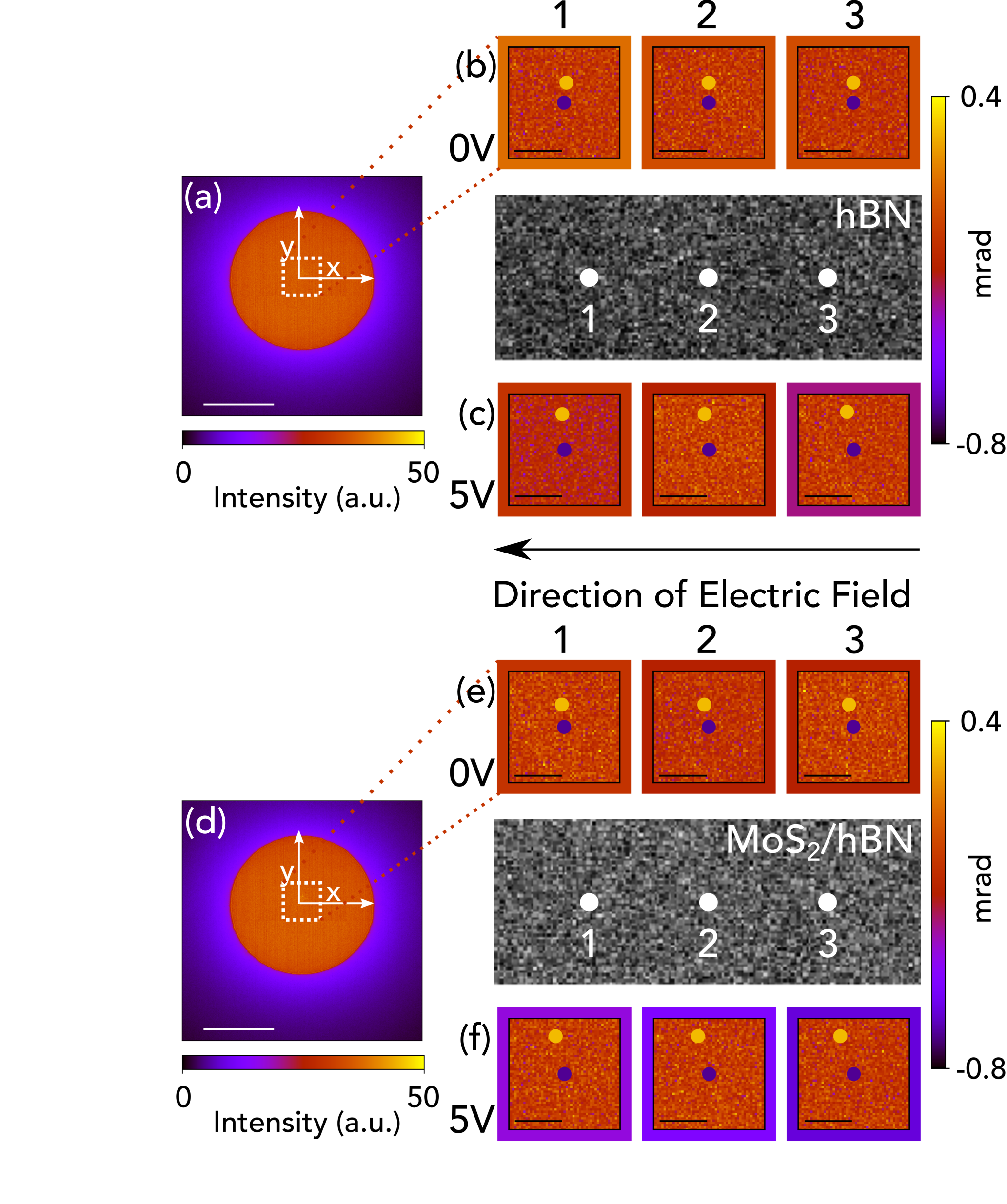}
\caption{Experimental scanning diffraction data acquired from the sample seen in center dark field TEM image. (a) CBED patterns captured at point 1 marked on the hBN sample. Colorbar intensity is in units of relative detection intensity. CBED patterns taken as a function of indicated positions when (b) 0V and (c) 5V are applied to the hBN sample. The purple dots refer to the center of mass of the CBED pattern taken from regions devoid of sample, while the yellow dots represent the center of mass of the CBED pattern taken from a particular position. The border color represents the degree of probe deflection in the x-direction.(d) CBED patterns captured at point 1 marked on the MoS$_2$/hBN sample. Colorbar intensity is in units of relative detection intensity. CBED patterns taken as a function of indicated positions when (e) 0V and (f) 5V are applied to the MoS$_2$/hBN sample. Scale bars represent 2 mrad. A clear shift in center of mass is observed when an external field is applied for both samples.}
\label{fgr:experimental}
\end{figure*}

Experimental results displaying \textit{I}\textsubscript{COM} values measured in a layer of hBN prior to applying and after applying external bias are presented in Figure \ref{fgr:experimental}. Prior to applying an external bias, the CBED pattern taken from spot 1 on the TEM image is provided in Figure \ref{fgr:experimental}a. In order to better observe the beam deflections present in these diffraction patterns, the center portion of individual CBED patterns is provided in Figure \ref{fgr:experimental}b. In these electron diffraction patterns, the green dot represents the \textit{I}\textsubscript{COM} of the vacuum probe that has not interacted with the sample. The yellow dots represent the \textit{I}\textsubscript{COM} of the CBED pattern taken from the specified positions of the hBN layer. The \textit{I}\textsubscript{COM} values are calculated using FM-STEM. When this measurement is repeated in the case where an external voltage of 5V is applied and an electric field is generated along the direction of the indicated arrow (Figure \ref{fgr:experimental}c), larger deflections in the \textit{I}\textsubscript{COM} values compared to the vacuum probe are detected. This suggests that this external stimulation leads to the incident electron probe experiencing additional momentum transfer.

To create a more electronically-relevant structure, a monolayer of MoS$_{2}$ is placed on top of the hBN layers. This semiconducting layer provides an alternative current pathway and this architecture represents a tunneling contact geometry previously explored for reducing contact resistance.\cite{RN962, RN1979} Once more, it is apparent that applying an external voltage leads to greater beam deflection in this geometry (Figure \ref{fgr:experimental}d-f).

\begin{figure*}
\includegraphics[width=\linewidth]{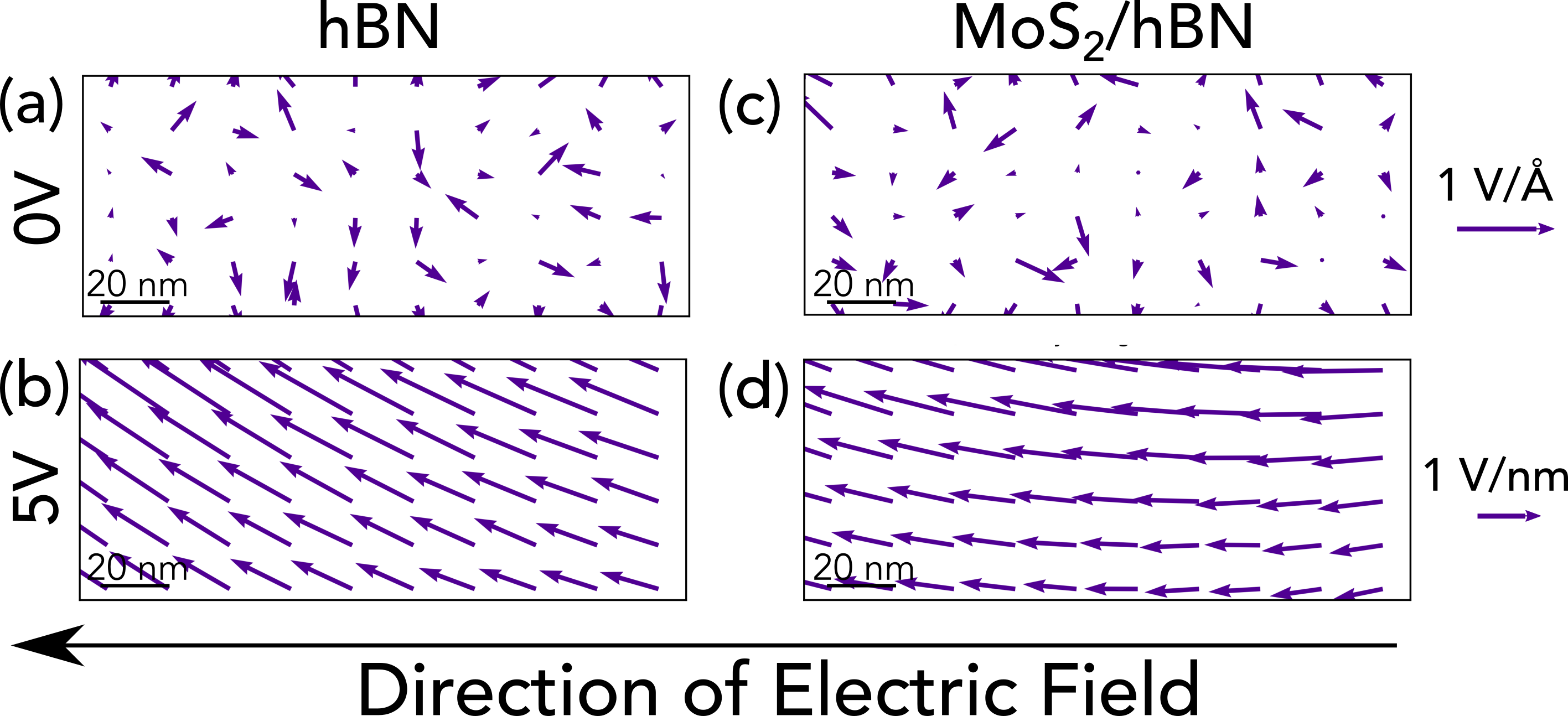}
\caption{Experimental electric field maps. In the absence of an applied  bias, electric field maps of the (a) hBN and (c) MoS$_2$/hBN samples are provided. Similar field maps in the case where an external bias of 5V is applied are provided in (b) and (d) for the hBN and MoS$_2$/hBN samples, respectively. In the case of an applied field, the measured electric field vectors align with the external potential gradient.}
\label{fgr:efield}
\end{figure*}

Eq. \ref{eqn:e-field} is applied to relate shifts in the \textit{I}\textsubscript{COM} of the CBED pattern to the electric field distribution. In the absence of an external electric field, these fields are seen in Figure \ref{fgr:efield}. The quasi-random pattern observed (Figure \ref{fgr:efield}a) likely represents electric fields arising from atomic orbitals that are captured by the uncorrected nanoscale electron probe.\cite{RN1751}

In order to calculate the electric field distribution when an external voltage of 5V is applied, electrons detected at each probe position are filtered using an annular, quadrant DPC virtual detector (Methods). Once again Eq. \ref{eqn:e-field} is used to calculate the electric field vector at each probe position. This map of the electric field distribution when an external voltage of 5V is applied is provided in Figure \ref{fgr:efield}b. This calculated field using the annular detector architecture agrees with our expectations for the uniform field distribution that would arise when an electrical potential is applied across hBN.

For the MoS$_2$/hBN sample, similar methods are used to characterize the resultant electrical fields in the absence of an external voltage and the long-range fields that arise when an external potential is applied. These, too, agree with prior expectations and are indicative of a uniform long-range field existing across the sample (Figure \ref{fgr:efield}c \& d).

Although based on the geometry of the system (5V applied across 1$\mu$m region), we expect the external field to have a magnitude on the order of 1 V/$\mu$m, the calculated field magnitude is found to be on the order of 1 V/nm. This discrepancy can potentially be explained by the fact that this measured electric field includes contributions from the electrostatic field in the sample plane as well as the leakage electric fields in the vacuum regions directly above and beneath the sample. We model the relative intensity of this leakage field by solving Laplace's equation for the sample geometry as seen in Figure S8. These calculations suggest that while the accumulated leakage field scales with sample field, they can persist for several hundred nanometers both above and beneath the sample. As such, rigorous analysis of the size of the electric field would be needed to quantitatively extract the precise electric field magnitudes within the sample, which is outside the scope of this study.

Finally, since Maxwell's theory states that the divergence of the electric field is proportional to the charge density (Methods), a local charge density distribution of this MoS$_2$/hBN region under applied electric field is provided in Figure S9. From this figure, it is clear that a carrier density roughly equivalent to 1-2 x $10^{13} cm^{-2}$ is observed when an external voltage is applied. These values are reasonable for monolayer MoS$_{2}$ and at such carrier densities, the material is expected to exist in a conductive regime with a 4 probe resistivity $<$ 100 k$\Omega$/sq.\cite{RN402}

In summary, we demonstrate the possibility for measuring the electrostatic field distribution that arises in 2D materials as a function of external bias. This technique involves the use of an annular DPC mask to relate shifts in the center of mass of the diffraction pattern as a function of position to local fields. Through simulated and experimental investigations, we determine that such masks provide a method for selectively probing the low spatial frequency information associated with long-range electric fields. This work demonstrates that these tools offer both a route to identifying non-uniformities in electric field as well as potential sources of noise or performance degradation during device operation in 2D architectures moving forward.

\section{ACKNOWLEDGEMENTS}

This material is based upon work supported by the National Science Foundation under Grant No. DMR-1929356. This work made use of the EPIC, Keck-II, and SPID facilities of Northwestern University’s NU\textit{ANCE} Center, which has received support from the SHyNE Resource (NSF ECCS-2025633), the IIN, and Northwestern's MRSEC program (NSF DMR-1720139). A.A.M. gratefully acknowledges support from the Ryan Fellowship and the IIN at Northwestern University. T.K.S. was supported by the Office of Naval Research (N00014-16-1-3055). P.L. was supported by Argonne National Laboratory. K.W. and T.T. acknowledge support from the Elemental Strategy Initiative conducted by the MEXT, Japan, Grant Number JPMXP0112101001, JSPS KAKENHI Grant Number JP20H00354 and the CREST (JPMJCR15F3), JST. The authors thank Erik Lenferink for assistance with sample characterization. Research reported in this publication was supported in part by instrumentation provided by the Office of The Director, National Institutes of Health of the National Institutes of Health under Award Number S10OD026871. The content is solely the responsibility of the authors and does not necessarily represent the official views of the National Institutes of Health. This research was supported in part through the computational resources and staff contributions provided for the Quest high performance computing facility at Northwestern University which is jointly supported by the Office of the Provost, the Office for Research, and Northwestern University Information Technology.

\section{AUTHOR CONTRIBUTIONS}
The manuscript was written through contributions of all authors. All authors have given approval to the final version of the manuscript.

\section{ADDITIONAL INFORMATION}
\textbf{Supplementary Information}
Supplementary Figs. 1–11.\\
\textbf{Competing Financial Interests:} The authors declare no competing financial interests.\\

\section{METHODS}
\renewcommand{\thesubsection}{\Alph{subsection}}
\renewcommand\thesubsubsection{\thesubsection.\arabic{subsubsection}}

\subsection{Material Synthesis}
Single crystal flakes of monolayer MoS$_{2}$ were grown through an atmospheric pressure chemical vapor deposition process (CVD). This process involved the use of 5 mg of MoO$_{3}$ (Sigma-Aldrich) powder, which is spread evenly throughout an alumina boat and capped by two face-down, 1 cm by 1 cm SiO$_{2}$/Si substrates (300 nm oxide). This boat is placed within a 1-inch diameter quartz tube at the center of a tube furnace. Similarly, 300 mg of sulfur pieces (Alfa-Aesar) were placed in another alumina boat located upstream and outside of the furnace. Following a 15 min purge of the sealed quartz tube with argon gas, the center of the furnace is heated to 700$^{\circ}$C over a period of 40 minutes and held at that temperature for 3 minutes. During this process, the flow rate of argon gas is maintained at 13 sccm. When the temperature of the furnace reached 575$^{\circ}$C during this ramp step, the alumina boat containing the sulfur is moved to a region inside the furnace where the temperature is approximately 150$^{\circ}$C. This is done with the help of a magnet. Following the growth process, the furnace is allowed to cool naturally.\\~\\

\subsubsection{DPC STEM Simulations}
Multislice simulations were performed using custom MATLAB\textsuperscript{\textregistered}  codes following the methods laid out by Kirkland.\cite{kirkland_2020} These simulations used similar microscope parameters as those used in acquiring the experimental datasets along with eight frozen phonon configurations. An external potential is applied by adding a mesh grid with linearly varying potentials from 0 to 10kV to the first layer of the multislice calculation. Traditional DPC images were acquired by processing the data using the virtual quadrant detectors seen in Figure \ref{fgr:simulation}. Electron signal in quadrants 1 and 3 were subtracted from one another to acquire DPC$_x$ images and electron signal in quadrants 2 and 4 were subtracted from one another to acquire DPC$_y$ images. For annular DPC images, the data is processed using the virtual annular quadrant detectors seen in Figure \ref{fgr:simulation} with an inner collection semi-angle of 28 mrad and an outer collection semi-angle of 32 mrad. DPC$_x$ and DPC$_y$ images were acquired through similar means.\\~\\

\subsubsection{Scanning Diffraction Image Processing}
4D STEM datasets were processed using custom Python scripts. For analysis associated with samples under no applied bias, the x and y components of the intensity center of mass at each probe position (\textit{I}\textsubscript{COM}) were calculated directly. For analysis associated with samples under an applied bias, transmitted electron signal both prior to applying and after applying applied bias is filtered using an annular quadrant detector. This virtual DPC detector is constructed to have an inner collection semi-angle of 28 mrad and an outer collection semi-angle of 32 mrad. In this case, DPC signal obtained under no applied bias is used as the background image and subsequently subtracted from the DPC signal obtained under applied bias to further eliminate the impact of short-range atomic potentials. The net signal in the x and y annular quadrants is used to approximate the x and y components of the \textit{I}\textsubscript{COM}. 

Following the protocol detailed by Fang et al. and Müller et al,\cite{RN1751, RN1812} electric field maps were calculated using the \textit{x} and \textit{y} components of the \textit{I}\textsubscript{COM} values. These values were related to the expectation value of the momentum transfer, $\left< p_{\bot }\right>$, through Eq. \ref{eqn:mom_transfer}. $\left< p_{\bot }\right>$ is calculated in units of eV s Å\textsuperscript{-1}. Eq. \ref{eqn:e-field} is used to relate the expectation value of the momentum transfer to electric field. $E_{\bot }$ is represented in the text using a quiver plot displaying both the $E_x$ and $E_y$ components and represented in units of V Å\textsuperscript{-1}.\\~\\

\noindent The total charge density ($\rho$), which is represented in units of elementary charge per Å\textsuperscript{2}, is calculated using Eq. \ref{eqn:charge_density}. In this equation,$\epsilon_0$ represents the vacuum permittivity. 

\begin{equation}
   div(E_{\bot }z)= \frac{\rho}{\epsilon_0} \label{eqn:charge_density}
\end{equation}

\subsubsection{Confocal Raman spectroscopy}
Raman spectra were obtained using a Horiba LabRAM HR Evolution Confocal Raman system under laser illumination with a wavelength of 532 nm and a power of 0.25mW.\\~\\

\printbibliography

\end{multicols} 

\newpage
\setcounter{figure}{0}
\setcounter{page}{1}
\setcounter{section}{0}

\renewcommand{\thepage}{S\arabic{page}}
\renewcommand{\thesection}{S\arabic{section}}
\renewcommand{\thefigure}{S\arabic{figure}}




\title{\vspace{-2cm}\fontsize{14}{16} \selectfont \noindent \textbf{Supplementary Information\\
Spatial Mapping of Electrostatics and Dynamics across 2D Heterostructures}}

\author{\fontsize{12}{15} \selectfont \noindent Akshay A. Murthy$^{1,2}$, Stephanie M. Ribet$^{1}$, Teodor K. Stanev$^3$,  Pufan Liu$^1$,\\
\fontsize{12}{15} \selectfont Kenji Watanabe$^4$, Takashi Taniguchi$^5$, Nathaniel P. Stern$^3$, Roberto dos Reis$^{1,6,*}$,\\
\fontsize{12}{15} \selectfont Vinayak P. Dravid$^{1,2,6,*}$}

\date{%
    \fontsize{10}{15} \selectfont \noindent
    $^1$Department of Materials Science and Engineering, Northwestern University, Evanston, IL, USA\\%
    $^2$International Institute of Nanotechnology, Northwestern University, Evanston, IL, USA\\
    $^3$Department of Physics and Astronomy, Northwestern University, Evanston, IL, USA\\
    $^4$Research Center for Functional Materials, National Institute for Materials Science, 1-1 Namiki, Tsukuba 305-0044, Japan\\
    $^5$International Center for Materials Nanoarchitectonics, National Institute for Materials Science, 1-1 Namiki, Tsukuba 305-0044, Japan\\
    $^6$The NU\textit{ANCE} Center, Northwestern University, Evanston, IL, USA\\~\\
    $^*$Corresponding Authors:\\
    Roberto dos Reis: \underline{roberto.reis@northwestern.edu}\\
    Vinayak P. Dravid: \underline{v-dravid@northwestern.edu}
}
\\~\\
\noindent KEYWORDS: \textit{In situ} electron microscopy, heterostructure, transition metal dichalcogenides, tunneling contacts, MoS$_{2}$, differential phase contrast
\\~\\
\noindent \textbf{Table of contents:\\
S1: Height profile of thin hBN support layers\\
S2: Spectroscopy and I-V results from prepared TEM sample\\
S3: Collapse of suspended TEM sample fabricated through alternative means\\
S4: hBN structure used for DPC simulations\\
S5: Comparison of inner collection semi-angles for annular DPC detector\\
S6: TEM diffraction pattern taken from MoS$_2$/hBN sample\\
S7: Variation in CBED with and without applied field\\
S8: Electric field simulation of TEM sample to understand impact of leakage field\\
S9: Charge Density across MoS$_2$/hBN}

\newpage
\title{\vspace{-2cm}\fontsize{14}{16} \selectfont \noindent \textbf{S1: Height profile of thin hBN support layers}}

\begin{figure}[hbt!]
\includegraphics[width=\linewidth]{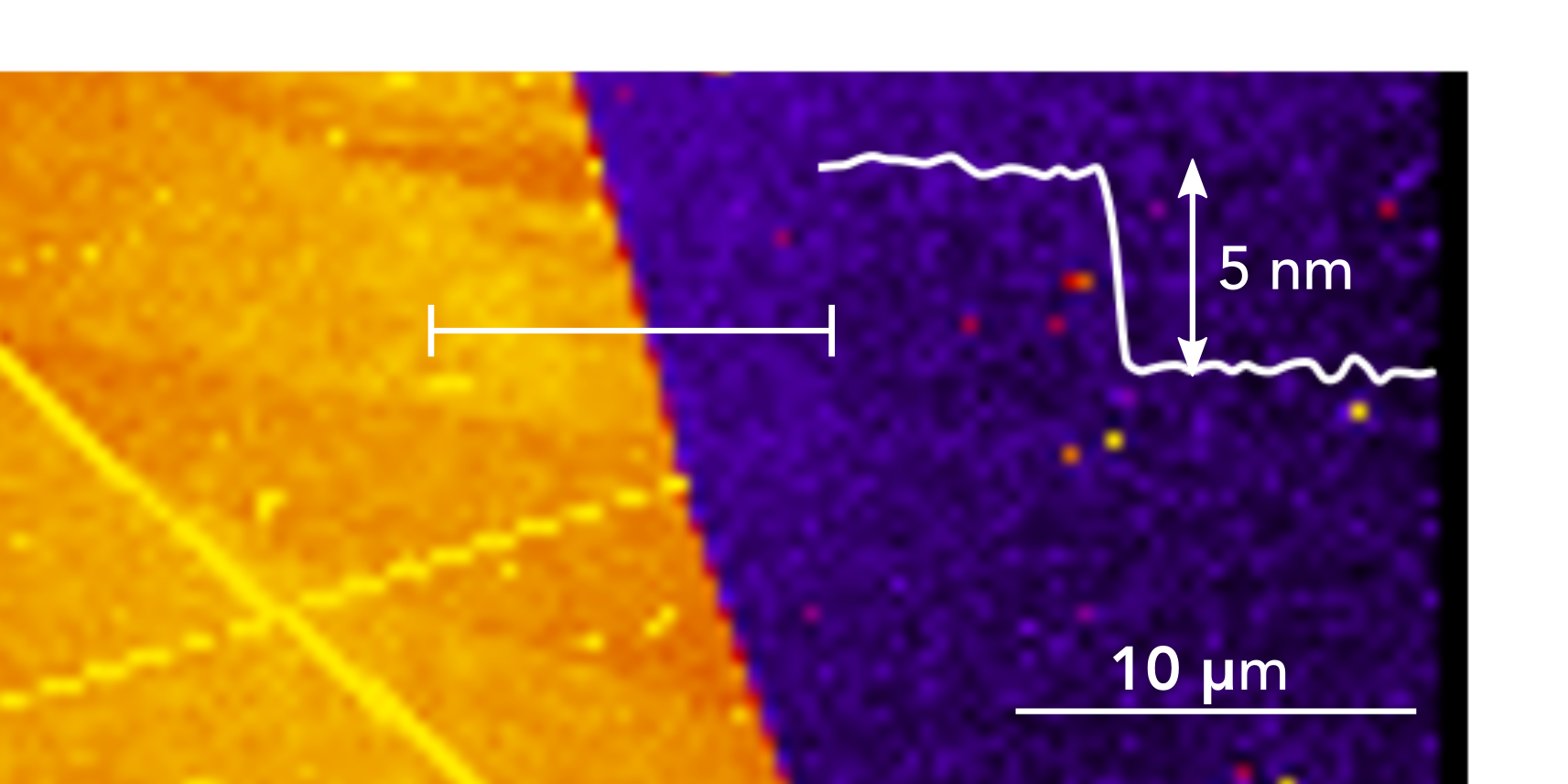}
\caption{Atomic force microscopy image taken from a flake of mechanically exfoliated hBN used as the support substrate. The flake thickness was found to be roughly 5 nm.}
\end{figure}

\newpage
\title{\vspace{-2cm}\fontsize{14}{16} \selectfont \noindent \textbf{S2: Spectroscopy and I-V results from prepared TEM sample}}

\begin{figure}[hbt!]
\includegraphics[width=\linewidth]{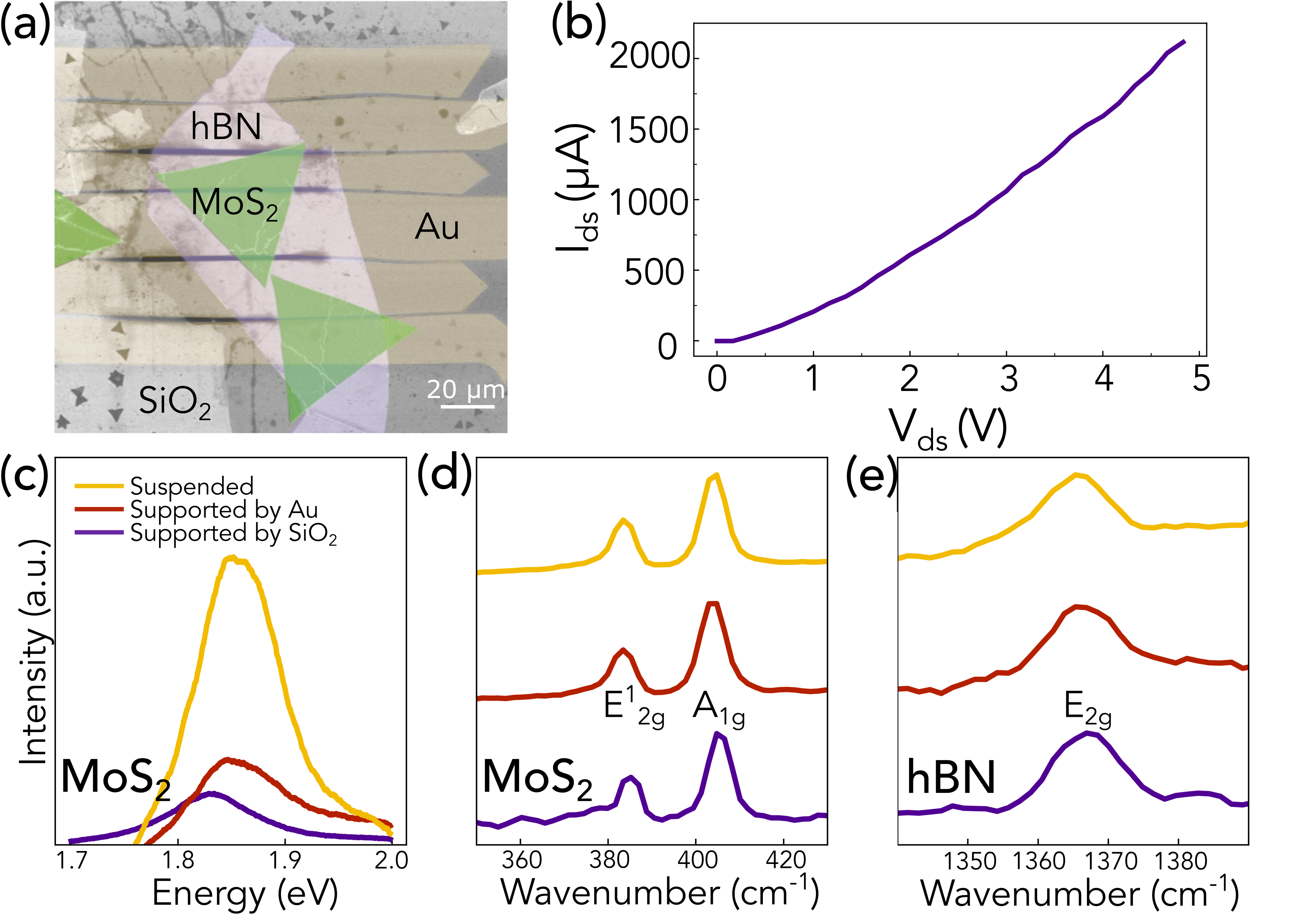}
\caption{(a) Colorized SEM image of suspended 2D heterostructure geometry. The gold region is indicative of gold electrodes. The purple region indicates the presence of h-BN, and the green region represents monolayer MoS$_{2}$, (b) Current (I$_{ds}$) - Voltage (V$_{ds}$) characteristics taken from sample indicate that an electric field can be applied across the sample, (c) Photoluminescence (PL) spectra taken from MoS$_{2}$/h-BN regions that are supported on SiO$_{2}$, supported on gold, and suspended. Raman spectrum taken from the same (d) MoS$_{2}$ and (e) h-BN regions are also provided. Due to the inherent differences in charge doping and strain, the position and magnitude of the PL peaks and the position of Raman modes vary with substrate.}
\end{figure}

\newpage
\title{\vspace{-2cm}\fontsize{14}{16} \selectfont \noindent \textbf{S3: Collapse of suspended TEM sample fabricated through alternative means}}

\begin{figure}[hbt!]
\includegraphics[width=\linewidth]{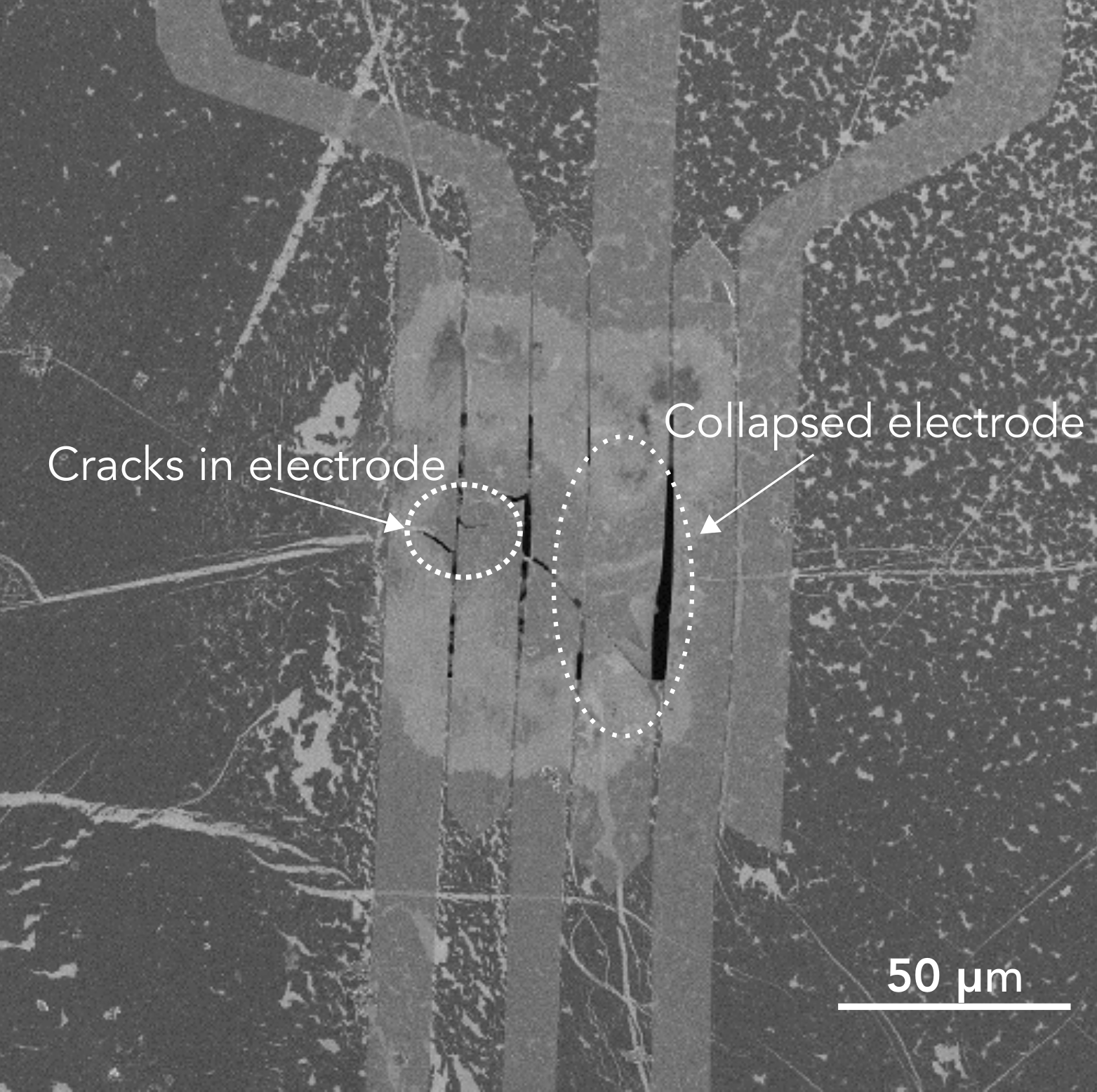}
\caption{Various challenges arise when trying to construct an \textit{in situ} TEM sample of a 2D heterostructure when using other methods. In this case, the electrodes were first transferred onto the TEM windows and 2D layers were sequentially transferred on top. Unfortunately, the repeated thermal processing involved in this setup prevents successful material suspension and creates cracking and fracturing of the metal electrodes.}
\end{figure}

\newpage
\title{\vspace{-2cm}\fontsize{14}{16} \selectfont \noindent \textbf{S4: hBN structure used for DPC simulations}}
\begin{figure}[hbt!]
\includegraphics[width=\linewidth]{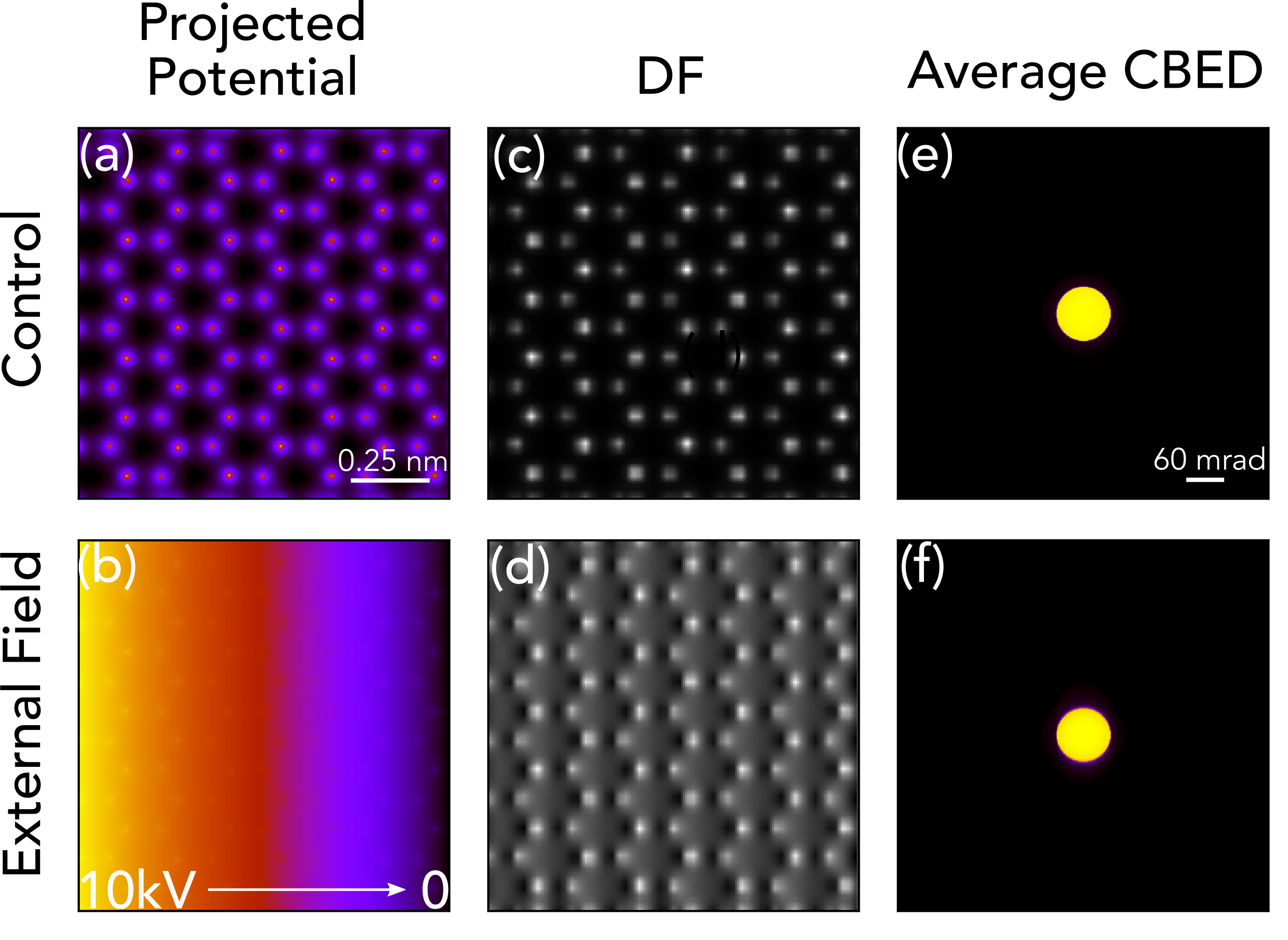}
\caption{(a) Projected potential of structure used for multislice simulation. (b) External potential applied to simulated structure. Simulated dark field images taken from structure with (c) and without (d) external potential. Average CBED taken from structure with (e) and without (f) external potential. The scale bar is consistent for the real space images in a-d as well as the diffraction space images in e-f}
\end{figure}

\newpage
\title{\vspace{-2cm}\fontsize{14}{16} \selectfont \noindent \textbf{S5: Comparison of inner collection semi-angles for annular DPC detector}}
\begin{figure}[hbt!]
\includegraphics[width=\linewidth]{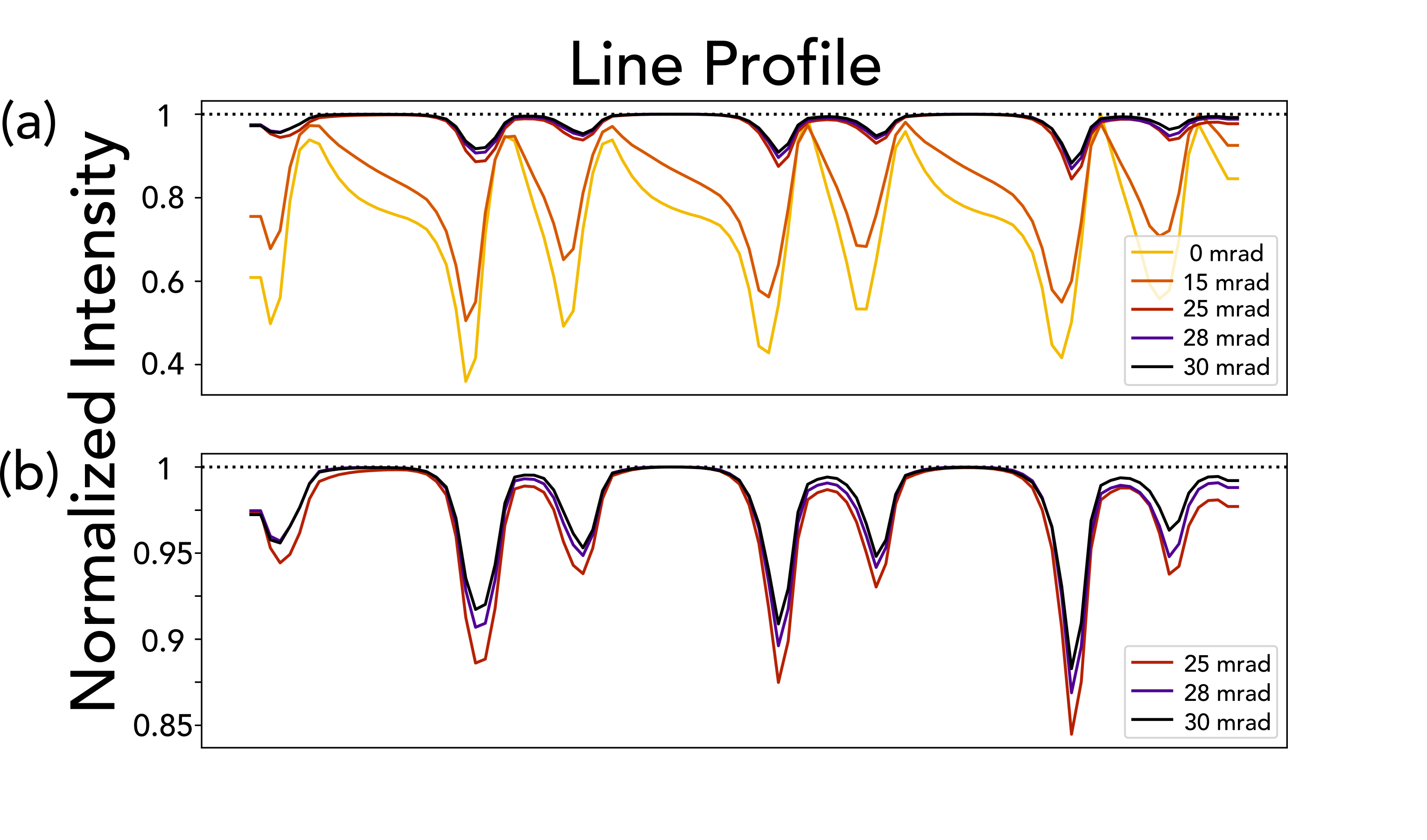}
\caption{(a) Normalized intensity line profiles taken from same region in Figure \ref{fgr:simulation} for inner collection semi-angles of 0 mrad, 15 mrad, 25 mrad, 28 mrad, and 30 mrad. (b) Normalized intensity lines profiles taken from same region in Figure \ref{fgr:simulation} replotted for inner collection semi-angles of 25 mrad, 28 mrad, and 30 mrad. Minimal intensity variation from atomic electrostatic fields is observed for both 28 mrad and 30 mrad inner collection angles. 28 mrad was selected to provide sufficient latitude in terms of beam and detector alignment in practical situations.}
\end{figure}

\newpage
\title{\vspace{-2cm}\fontsize{14}{16} \selectfont \noindent \textbf{S6: TEM diffraction pattern taken from sample}}
\begin{figure}[hbt!]
\includegraphics[width=\linewidth]{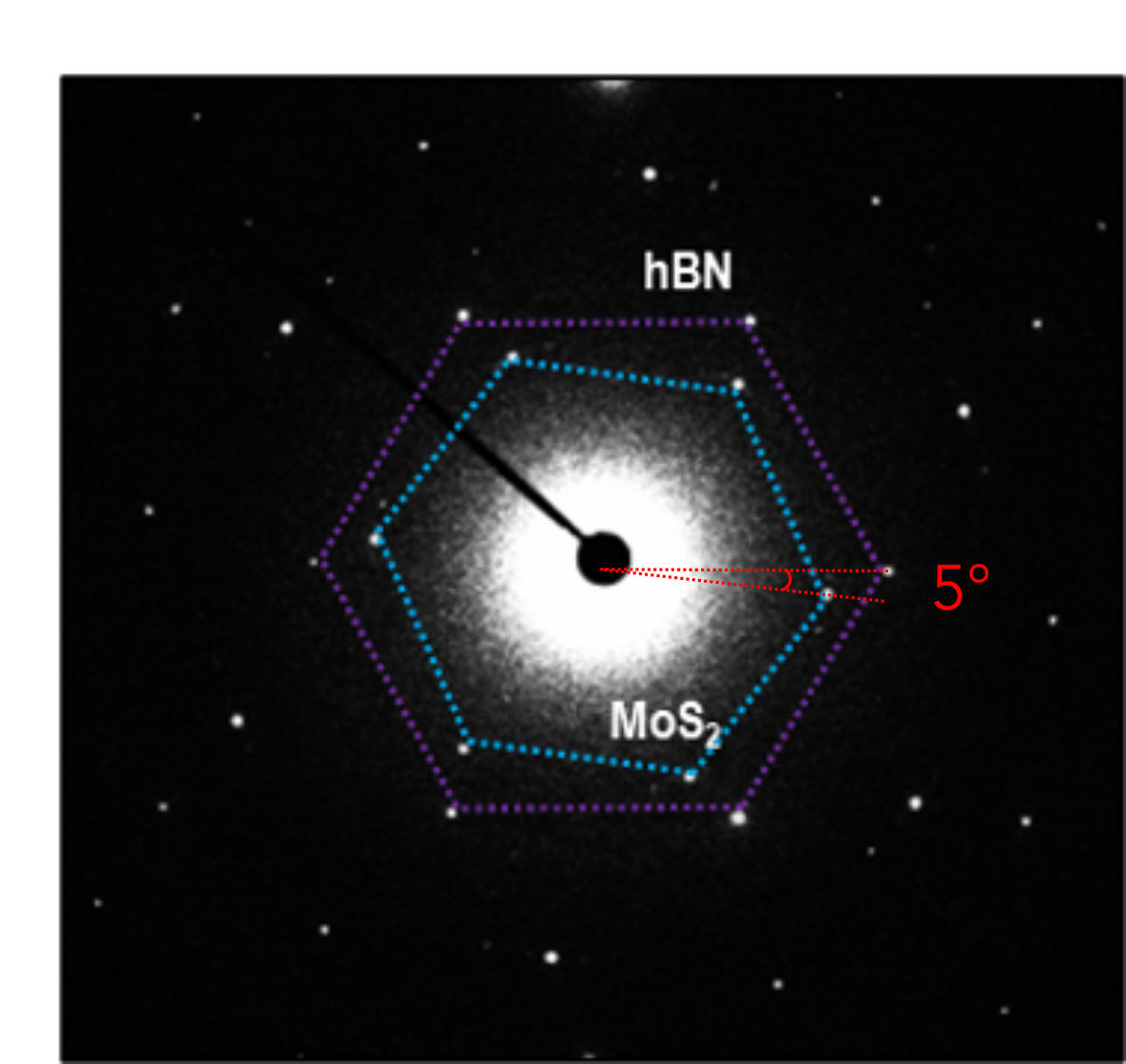}
\caption{TEM selected area diffraction pattern taken from MoS$_2$/hBN sample indicates that the two layers are present and offset by 5$^{\circ}$.}
\end{figure}

\newpage
\title{\vspace{-2cm}\fontsize{14}{16} \selectfont \noindent \textbf{S7: Variation in CBED with and without applied field}}
\begin{figure}[hbt!]
\includegraphics[width=80mm]{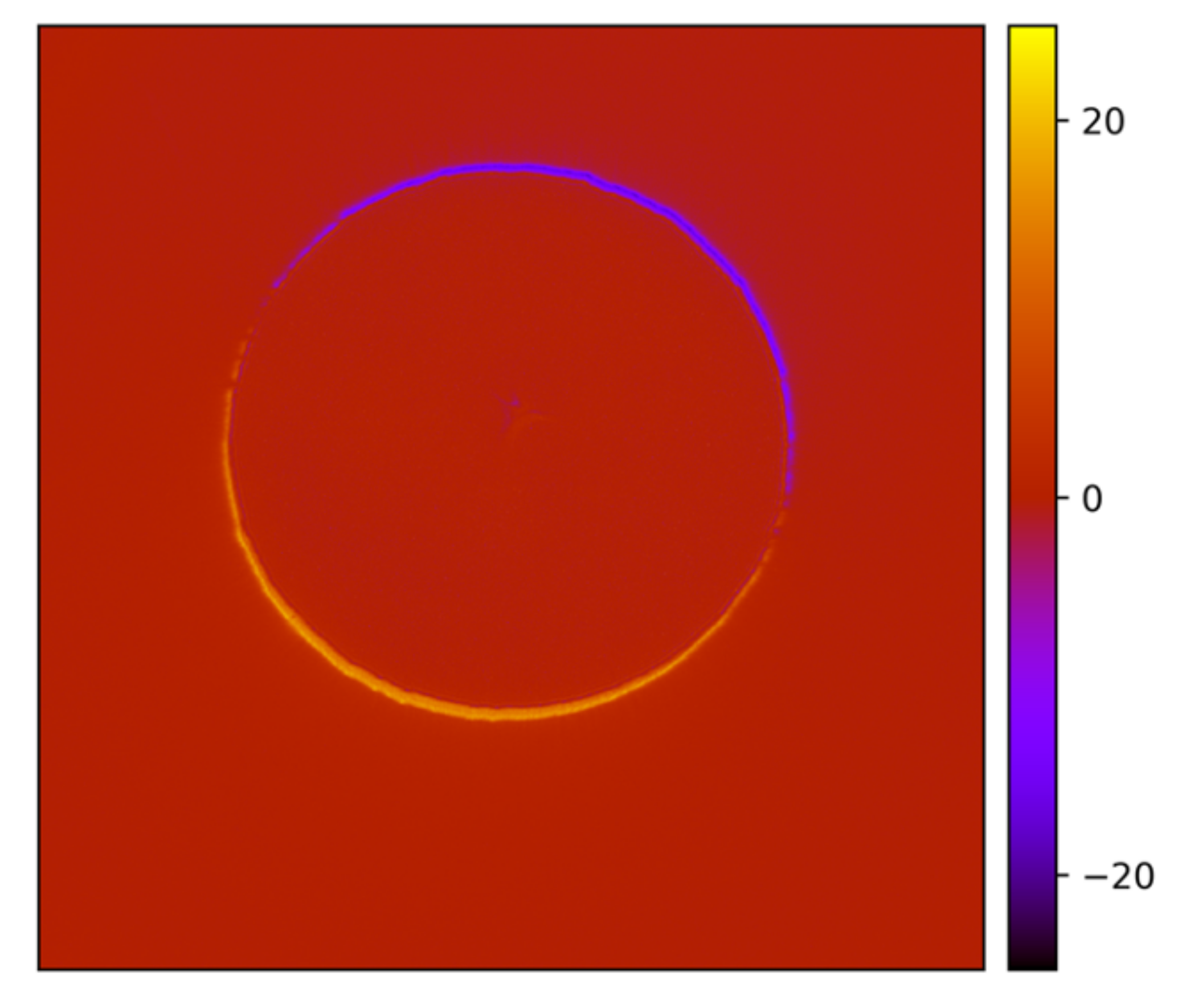}
\caption{Difference in experimental CBED patterns taken from MoS$_2$/hBN region at position 1 on Figure \ref{fgr:experimental} between V$_{ds}$=0V and V$_{ds}$=5V conditions. This colormap indicates that when the probe is smaller than the feature, the \textit{I}\textsubscript{COM} signal is measured from the shifts of the unscattered beam as opposed to asymmetries within the disc.}
\end{figure}

\newpage
\title{\vspace{-2cm}\fontsize{14}{16} \selectfont \noindent \textbf{S8: Electric field simulation of TEM sample to understand impact of leakage field}}
\begin{figure}[hbt!]
\includegraphics[width=80mm]{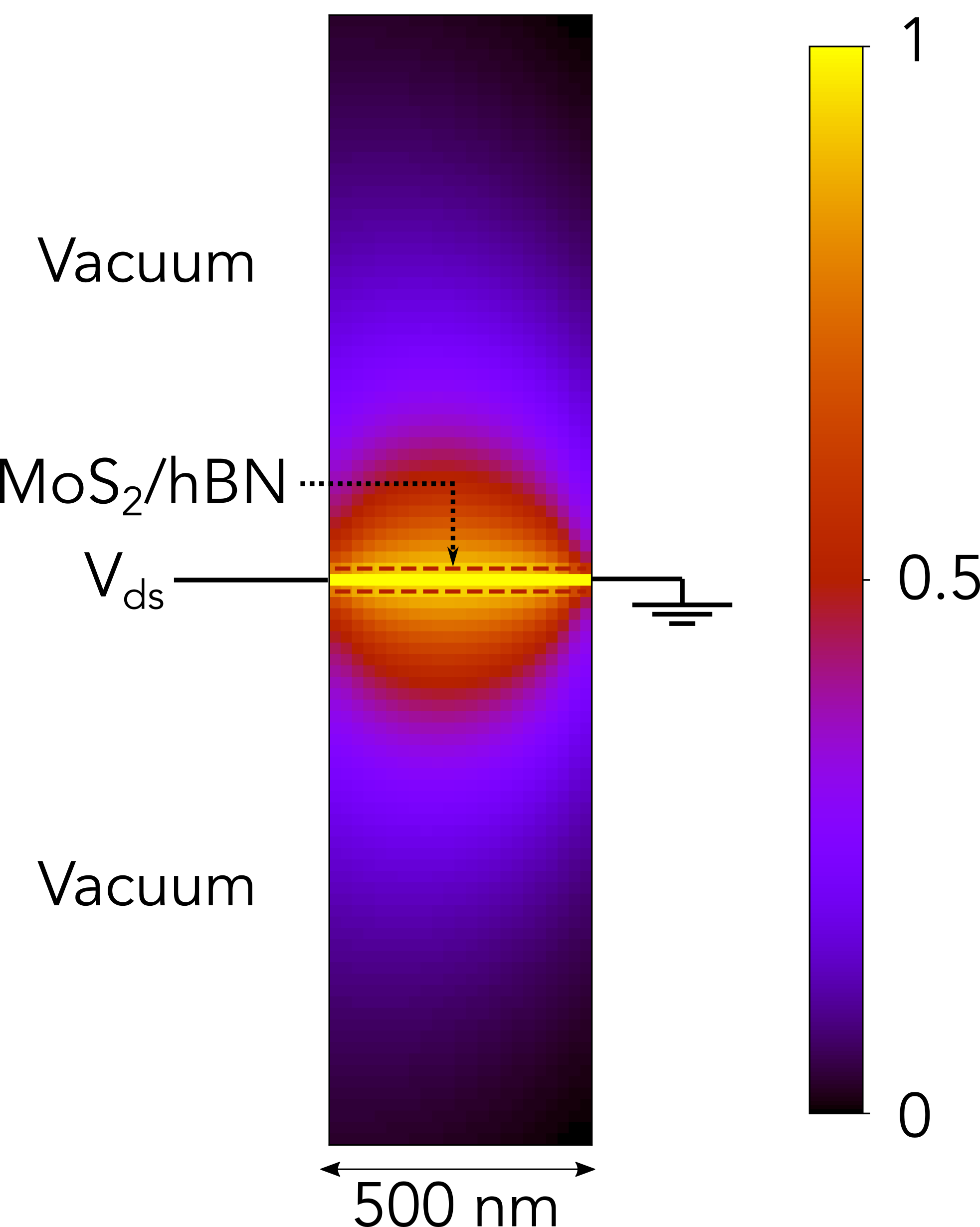}
\caption{Simulated relative electric field intensities in the sample as well as a 1$\mu$m vacuum region directly above and below the sample plane. For this simulation, boundary condition values of 0V at the ground electrode and a positive voltage at the source electrode (V\textsubscript{ds)} were used. The Laplace equation was solved to determine the electric potential at each finite location. By taking the in-plane spatial gradient of these values, the in-plane electric field distribution was calculated.}
\end{figure}

\newpage
\title{\vspace{-2cm}\fontsize{14}{16} \selectfont \noindent \textbf{S9: Charge Density across MoS$_2$/hBN}}
\begin{figure}[hbt!]
\includegraphics[width=\linewidth]{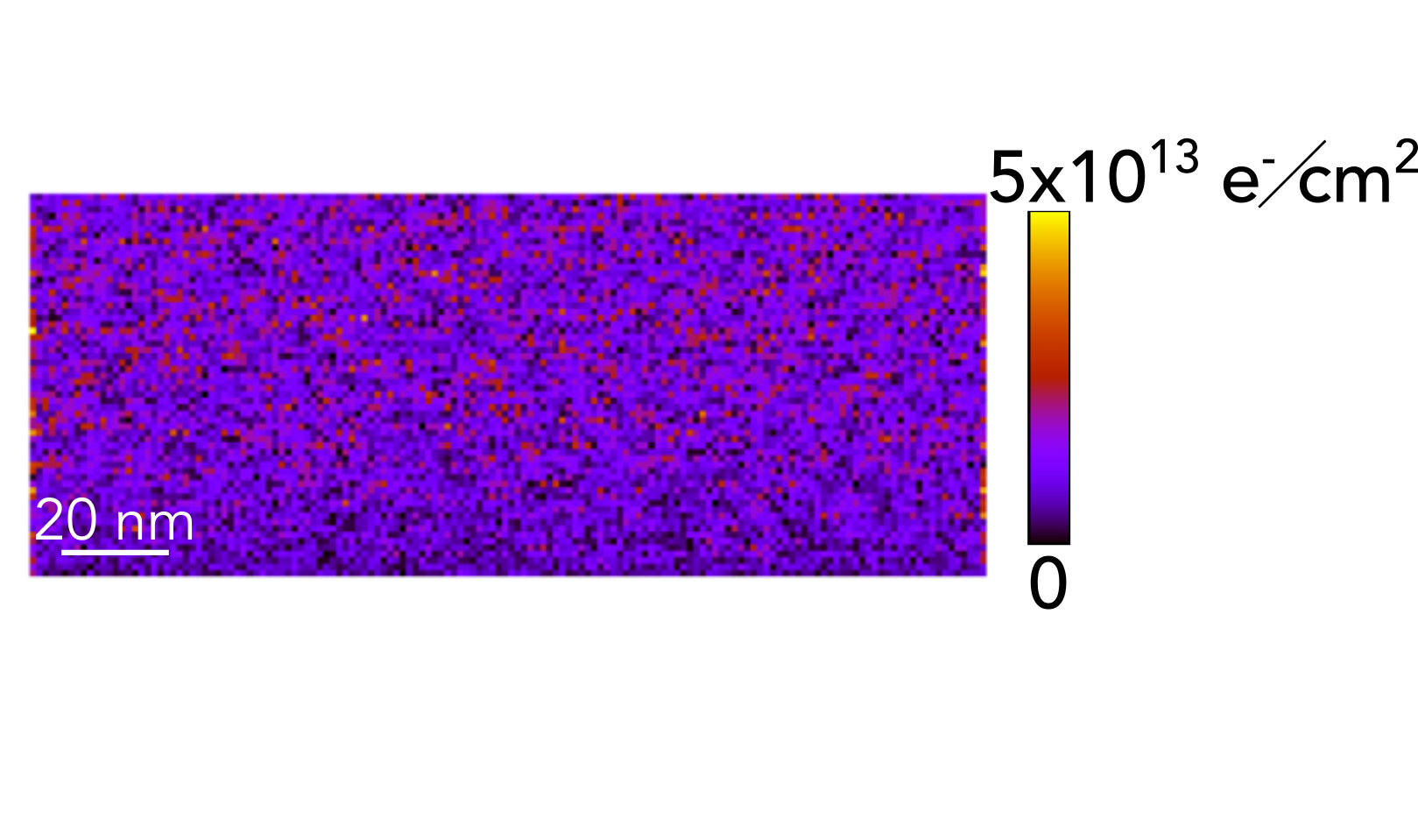}
\caption{Charge density map across MoS$_2$/hBN region calculated from electric field map in Figure \ref{fgr:efield}d.}
\end{figure}

\end{document}